\documentstyle[12pt]{article}
\begin{document}
\def\mbs{\mbox{\boldmath$\sigma$}}
\def\Ket#1{||#1 \rangle}
\def\Bra#1{\langle #1||}
\def\sss{\scriptscriptstyle}
\def\ss{\scriptstyle}
\def\endauthors{}
\def\authors#1\endauthors{#1}
\def\intsum {\int\!\!\!\!\!\!\!{\ss \sum}}
\def\intsums {\int\!\!\!\!\!{\ss\Sigma}}
\def\be{\begin{equation}}
\def\ee{\end{equation}}
\def\br{\begin{eqnarray}}
\def\er{\end{eqnarray}}
\def\brn{\begin{eqnarray*}}
\def\ern{\end{eqnarray*}}
\def\rf#1{{(\ref{#1})}}
\def\ket#1{|#1 \rangle}
\def\bra#1{\langle #1|}
\def\Ket#1{||#1 \rangle}
\def\Bra#1{\langle #1||}
\begin{titlepage}
\pagestyle{empty}
\baselineskip=21pt
\vskip .2in
\begin{center}
{\large{\bf
Competition between standard and exotic double beta decays}}
\end{center}
\vskip .1in
\authors
\centerline{C.~Barbero${}^{\dagger}$, F.~Krmpoti\'{c}${}^{\dagger}$,
A.~Mariano${}^{\dagger}$}
\vskip .15in
\centerline{\it Departamento de F\'\i sica, Facultad de Ciencias}
\centerline{\it
Universidad Nacional de La Plata, C. C. 67, 1900 La Plata, Argentina.}
\vskip .1in
\endauthors
\vskip 0.5in
\centerline{ {\bf Abstract} }
\baselineskip=18pt
\bigskip
\vspace{0.5in}
\noindent
We discuss the contributions of higher order terms in weak
Hamiltonian to the standard two-neutrino double beta decay.
The formalism for the unique first forbidden transitions has been developed,
and it is shown that they can alter the two-electron energy spectrum.
Yet, their effect is too small to screen the detection of exotic neutrinoless
double beta decays, which are candidates for testing the physics beyond the
standard model.

\vspace{0.5in}


\vspace{0.5in}

$^{\dagger}$Fellow of the CONICET from Argentina.
\end{titlepage}

The double beta ($\beta\beta$) experiments furnish a unique window onto
whatever new physics may replace the Standard Electroweak Model (SM).
To be observable the new
physics should: i) violate the electron-lepton-number $(L_e)$ conservation,
that is fulfilled in the SM, and/or ii) fit
scalar particles (currently called Majorons), light enough to be produced
in the $\beta\beta$ decay.

The quantity that is used to discern experimentally between the ordinary
SM two-neutrino decays ($\beta\beta_{2\nu}$) and the exotic neutrinoless
$\beta\beta$ events, both without ($\beta\beta_{0\nu}$) and with Majoron
emissions ($\beta\beta_{\sss M}$), is the electron energy spectrum
$d\Gamma/d\epsilon$ of the decay rate $\Gamma$, as a function of the sum
$\epsilon= \epsilon_1+\epsilon_2$ of the energies of the two emitted
electrons.
The $\beta\beta_{2\nu}$ decay exhibit a continuous spectra in the interval
$2\le\epsilon\le Q$, where $Q=E_{\sss I}-E_{\sss F}$ is the released
energy.
On the other hand, when no new light particles are created, the
$L_e$-violating terms in the weak Lagrangian, that generate a
Majorana mass for the  neutrino, can be identified if they produce
the $\beta\beta_{0\nu}$ decay, with the energy spectrum that is just
a spark at the energy $Q$.
The spectra for the $\beta\beta_{\sss M}$ decays are also continuous, but
their shapes are clearly differentiated  from that for the $\beta\beta_{2\nu}$
decay, and depend on whether one or two Majorons are emitted and on the
leptonic charge ($L_e=0,-1,-2$) they carry \cite{Bur93,Car93,Bam95}.
Thus, both $\beta\beta_{0\nu}$ and $\beta\beta_{\sss M}$ processes, that are
potentially capable to reveal the new physics, are clearly distinguishable
from the SM $\beta\beta_{2\nu}$ decay.

The sensitivity of the $\beta\beta$ decay experiments is steadily and
constantly increasing.
For instance, while  the pioneer laboratory measurement of the
$\beta\beta_{2\nu}$ decay in $^{82}Se$ has relied on
only 40 events \cite{Ell87}, the most recent experiment on $^{76}Ge$
\cite{Kla98} was done with high statistics ($\sim 20000$ counts).
Another example is the evolution of the half life limit for the
$\beta\beta_{0\nu}$ decay in $^{76}Ge$.
>From the first measurement in 1952, $T_{0\nu}>2\cdot 10^{16}$ y
\cite{Fre52}, it has varied to $T_{0\nu}>2.2\cdot 10^{22}$ y in 1983
\cite{Avi83}, while the most recent value is $T_{0\nu}>1.2\cdot 10^{25}$
y \cite{Kla98}.
By comparing the last one, as well as the  measured half life limit for
$\beta\beta_{\sss M}$ decay: $T_{\sss M}>1.67\cdot 10^{22}$ y, with
the corresponding half life for $\beta\beta_{2\nu}$ decay:
$T_{2\nu}\cong 1.77\cdot 10^{21}$ y, it can be said that presently
are being observed effects of the order of $10^{-4}$ at $\epsilon \sim Q$
and of the order of $10^{-1}$ at $\epsilon \sim Q/2$.
There are also several ongoing and planned experiments that are
supposed to allow for measurements of still smaller effects.
The most promising one seems the GENIUS project \cite{Kla98}, which
is supposed to test the $\beta\beta_{0\nu}$ half-life of
$^{76}Ge$ up to a limit of $T_{0\nu}>5.7\cdot 10^{28}$ y.
One might expect that the sensitivity for probing the $\beta\beta_{\sss M}$
decay will be improved accordingly as well.

The measured limits on the exotic $\beta\beta$ transition probabilities
are being rapidly translated into more stringent constrains on the
parameters of new theoretical developments in particle physics, such as:
Majorana mass of the light neutrinos, right-handed weak couplings,
right-handed weak coupling involving heavy Majorana neutrinos, massless
Majoron emission, R-parity breaking in the supersymmetric model, {\it etc.}
They have also broad consequences on the history of the primordial
universe, evolution  of stellar object and astrophysics of supernovas.

In confronting the experimental data with theory, the validity of allowed
approximation (A) is usually assumed for the standard $\beta\beta_{2\nu}$
decay.
This implies to consider only virtual states with spin and parity
$J^\pi=0^+$ and $1^+$, which contribute via the nuclear operators
$g_{\sss V}\tau^+$ and $g_{\sss A}\tau^+\mbs$, respectively.
The higher order effects, coming from the parity-forbidden (PF) virtual states
with
$J^\pi= 0^-,1^-,2^-$, have been ignoring almost entirely all along by
workers in the field, simply because they expected them to be small.
But, in planing future experiments and in searching for exotic decays,
it might be important to know how small  these effects are, and whether
they could eventually lead to experimental consequences similar to that
produced by the former.

Besides, in the charged Majoron (CM) model designed by Burgess and Cline
\cite{Bur93} and by Carone \cite{Car93}, which is the most hopeful one
to be observed experimentally among the new Majoron models \cite{Hir96},
the $\beta\beta$ decay proceeds via relativistic corrections in the hadronic
current. More, the nuclear matrix element is  of the
form ${\cal M}_{\sss CM}=  {\cal M}^+_{\sss CM} - {\cal M}^-_{\sss CM}$,
with ${\cal M}_{\sss CM}^{\pm}$ being the contributions of two heavy
Dirac neutrinos with masses $M_{\pm}$,
and there is a strong destructive interference between
${\cal M}^+_{\sss CM}$ and ${\cal M}^-_{\sss CM}$ when $M_+\cong M_-$
\cite{Bar96}.
Therefore it might be also interesting to compare the outcome of the CM model
with that arising from the PF transitions in the standard $\beta\beta$ decay.
\footnote{In fact, except for the neutrino mass term, the
$\beta\beta_{0\nu}$ decay also arises from the higher order effects
\cite{Bar98}.}

Naively thinking it could be inferred that the PF contributions to the
$\beta\beta_{2\nu}$ amplitude are of the order of $(RQ/4)^2$, where $R$ is the
nuclear radius and the factor $4$ comes from the fact that the $Q$-value is
shared
by $4$ leptons. Therefore, as for medium heavy nuclei $(RQ/4)^2\sim 10^{-3}$,
they would alter the half lives and spectrum shapes at the level of $10^{-3}$
or $10^{-6}$, depending on  whether there are interference or not between the
A and PF contributions, {\it i.e.} whether the PF matrix elements enter
linearly or only quadratically in the expression for the decay rate.

The above estimate, however,
it is not valid for the
non-unique (NU) transitions  with  $\Delta J^\pi=0^-,1^-$, due to:
i) the electron $s$-waves contributions via the velocity dependent terms in the
hadronic current, and
ii) the Coulomb enhancement of the matrix elements coming from the $p_{1/2}$
waves.
When these effects are considered,
and because the interference between the NU and A matrix elements,
the theoretical half lives are increased in $\sim 30\%$ \cite{Wil88,Bar95}.
Yet, due to the nuclear structure uncertainties, we cannot disentangle from
the measured half lives alone how large is the effect of the NU matrix
elements, even when the experimental errors are relatively small.

We have also pointed out \cite{Bar95} that
the NU transitions
do not modify the allowed shape of the two-electron spectrum, and
that the unique (U) transitions states should be examined.
It should be remembered that the spectrum shape of a single $\beta$ transition
of this type, provides for the emission of more high- and low-energy
electrons, than are found in spectra that have the allowed shape 
\cite{Eva55,Beh82}.
Thus, it sounds appropriate to speculate about a similar effect in the
$\beta\beta$ decay. Simultaneously,
it is essential to find out whether the U matrix
elements contribute
linearly or quadratically to the
$\beta\beta_{2\nu}$ decay rate.

Meanwhile, Civitarese and Suhonen \cite{Civ96}, have pointed out
that the effect of the U transitions on the $T_{2\nu}$ in $^{76}Ge$
can be disregarded, but they have not
discussed  at all the corresponding  spectrum shape.
Without performing any theoretical development, it is
simply assumed that the U matrix element
come through quadratically, and that the ratio between the phase-space
factors, for A and U
$\beta\beta_{2\nu}$ transitions, is of the order of $10^6$.
Besides, in the same work it is suggested that the NU transitions should be
retarded as well by the factor $(RQ/4)^4\sim 10^{-6}$.

Before presenting the numerical results
for the $\beta\beta_{2\nu}$ decay
through the virtual states $J^\pi=2^-$, we briefly sketch
the derivation of the corresponding
formulae, which has not been done so far.
The
$\beta\beta_{2\nu}$ decay rate reads
\be
d\Gamma_{2\nu}=2\pi \intsum |R_{2\nu}|^2 \delta(\epsilon_1+\epsilon_2
+\omega _1+\omega _2-Q)\prod_{k=1}^2d{\bf p}_k d{\bf q}_k,
\label{1}\ee
where the symbol ${\intsums}$ represents both the summation on lepton spins, 
 and the integration on neutrino momenta and electron directions.
For a transition from the initial state $\ket{0_{\sss I}}$ in the $(N,Z)$
nucleus to
the final state $\ket{0_{\sss F}}$ in the $(N-2,Z+2)$ nucleus
(with energies $E_{\sss I}$ and $E_{\sss F}$ and spins and parities
$J^{\pi}=0^+$) the
transition amplitude is evaluated via the second order Fermi golden rule:
\br
R_{2\nu}&=&\frac{1}{2(2\pi)^6}\sum_{\ss N}[1-P(e_1e_2)][1-P(\nu_1\nu_2)]
\frac{\bra{0_{\sss F}^+}H_{\sss W}(e_2\nu_2)
\ket{{\ss N}}\bra{{\ss N}}H_{\sss W}(e_1\nu_1)
\ket{0_{\sss I}^+}}{E_{\sss N}-E_{\sss I}+\epsilon_1+\omega _1},
\label{2}\er
where $e_i\equiv (\epsilon_i,{\bf p}_i,s_{e_i})$, $\nu_i\equiv 
(\omega _i,{\bf q}_i,s_{\nu_i})$, $P(l_1l_2)$
exchanges the quantum numbers of leptons $l_1$ and $l_2$, and ${\ss N}$
runs over all levels in the $(N-1,Z+1)$ nucleus.
The weak Hamiltonian reads
\be
H_{\sss W}(e\nu)=\frac{G}{\sqrt{2}}\int d{\bf x} j_\mu  ({\bf x})
J^{\mu  \dagger}({\bf x})+h.c.,
\label{3}\ee
where  $G=(2.996\pm 0.002){\times} 10^{-12}$ is the
Fermi coupling constant (in natural units),
$j^{\mu}({\bf x})$
is the usual left-handed leptonic current \cite{Doi85}, and for
the hadronic current
\begin{eqnarray}
J^\mu ({\bf x})&=&\left(\rho_{{\sss {V}}}({\bf x})-
\rho_{{\sss {A}}}({\bf x}),
{\bf j}_{{\sss {V}}}({\bf x})-{\bf j}_{{\sss {A}}}({\bf x})\right),
\label{4}\end{eqnarray}
the following non-relativistic
approximation will be used \cite{Bar98}
\begin{eqnarray}
\rho_{{\sss {V}}}({\bf x})&=&g_{{\sss {V}}} \sum_n\tau_n^+ 
\delta({\bf x}-{\bf r}_n),\nonumber\\
\rho_{{\sss {A}}}({\bf x})&=&\frac{g_{{\sss {A}}}}{2M_{\sss N}}
\sum_n\tau_n^+[\mbs_n\cdot{\bf p}_n\delta({\bf x}-{\bf r}_n)
+\delta({\bf x}-{\bf r}_n) \mbs_n\cdot{\bf p}_n] ,\nonumber\\
{\bf j}_{{\sss {V}}}({\bf x})&=&\frac{g_{{\sss {V}}}}{2M_{\sss N}}
\sum_n\tau_n^+[{\bf p}_n\delta({\bf x}-{\bf r}_n)
+\delta({\bf x}-{\bf r}_n){\bf p}_n
+f_{\sss W}{\mbox{\boldmath$\nabla$}}{\times}\mbs_n
\delta({\bf x}-{\bf r}_n)]  ,\nonumber\\
{\bf j}_{{\sss {A}}}({\bf x})&=&g_{{\sss {A}}}\sum_n\tau_n^+ 
\mbs_n\delta({\bf x}-{\bf r}_n),
\label{5}\end{eqnarray}
where
$M_{\sss N}$ is nucleon mass, and $g_{\sss V}$, $g_{\sss A}$ and
$f_{\sss W}$ are, respectively, the vector, axial-vector and
weak-magnetism effective coupling constants.

In the discussion of the $\beta\beta_{2\nu}$ decay we ignore both the
weak-magnetism term, and the action of the velocity dependent terms on
the lepton current.
These terms cause the "second-forbidden" contributions, which do not
alter the electron spectrum shape and will be discussed elsewhere.
Additionally, it will be assumed that the Coulomb energy of the electron
at the nuclear radius is larger than its total energy, which leads to
the  $\xi$-approximation \cite{Bar95,Beh82}.
Thus, for the purposes of the present study, and after a
lengthy algebra, we
cast the weak Hamiltonian in a rather novel form
\be
H_{\sss W}(e\nu)=-\frac{G}{2}\sum_{\pi J}
{\sf W}^\pi_J\cdot{\sf L}_J(e\nu ),
\label{6}\ee
which can be used for multiple purposes. Here ${\sf W}^+_J$ and
${\sf W}^-_J$
are, respectively, the allowed and forbidden nuclear operators, and

\be
{\sf L}_J(e\nu )=sg(s_\nu)
\sqrt{\frac{\epsilon+1}{2\epsilon} F_0(\epsilon)}
\chi^\dagger(s_e)\left(1-\frac{\mbs\cdot{\bf p}}{\epsilon+1}\right)\ell_J
(1-\mbs\cdot\hat{\bf q})\chi(-s_\nu),
\label{7}\ee
are the leptonic matrix elements, with $\chi(s)$ being the usual Pauli spinor.
The leptonic operators $\ell_J$ are listed in Table \ref{tab1}, together
with ${\sf W}^\pi_J$.

\begin{table}[h]
\begin{center}
\caption {Operators $\ell_J$ and ${\sf W}^\pi_J$ for
different multipoles $J$;
$\bar {p}=p[F_1(\epsilon)/F_0(\epsilon)]^{1/2}$, ${\bf v}={\bf p}/M_{\sss N}$
and $\xi=\alpha Z/2R$.}
\label{tab1}
\begin{tabular}{cccc}
\hline
$J$&$\ell_J$&${\sf W}^+_J$& ${\sf W}^-_J$\\
\hline
$0$&$1$&$g_{\sss V}$&$-g_{\sss A}(\mbs\cdot{\bf v}+\xi i\mbs\cdot{\bf r})$\\
$1$&$\mbs$&$g_{\sss A}\mbs$&
$-g_{\sss V}{\bf v}-\xi[g_{\sss V}i{\bf r}-g_{\sss A}
(\mbs{\times}{\bf r})]$\\
$2$&$\left[\mbs\otimes\left({\bf q}
+\bar{{\bf p}}\right) \right]_2$&-
&$ig_{\sss A}(\mbs\otimes{\bf r})_2/\sqrt{5}$\\
\hline\end{tabular}\end{center}
\end{table}

In the next step we evaluate the transition amplitude and get
\be
R_{2\nu}=\frac{G^2}{4(2\pi)^6}[1-P(\nu_1\nu_2)]\left[\frac{}{}
{\sf L}_{0}(e_1\nu_1)\cdot{\sf L}_{0}(e_2\nu_2)
\left({\cal M}_{2\nu}^{\sss A}+{\cal M}_{2\nu}^{\sss NU}\right)
-{\sf L}_{2}(e_1\nu_1)\cdot{\sf L}_{2}(e_2\nu_2)
{\cal M}_{2\nu}^{\sss U}\right],
\label{8}\ee
where ${\cal M}_{2\nu}^{\sss A}={\cal M}_{2\nu}(0^+) +{\cal M}_{2\nu}(1^+)$,
${\cal M}_{2\nu}^{\sss NU}={\cal M}_{2\nu}(0^-) +{\cal M}_{2\nu}(1^-)$,
and ${\cal M}_{2\nu}^{\sss U} \equiv{\cal M}_{2\nu}(2^-)$
are, respectively, the
 $\beta\beta_{2\nu}$ matrix elements
for the  A, NU and U transitions, and
\be
{\cal M}_{2\nu}(J^\pi)
=\sum_{\alpha}(-1)^J\frac{
\Bra{0^+_{\sss F}}{\sf W}^\pi_J\Ket{J^\pi_{\alpha}}\Bra{J^\pi_{\alpha}}
{\sf W}^\pi_J\Ket{0^+_{\sss I}}}{E_{J^\pi_\alpha}-E_{0^+_{\sss I}}+Q/2}.
\label{9}\ee

After introducing \rf{8} into \rf{1}
and performing  the spin summations and angular integration,
the contribution of the lepton matrix elements
${\sf L}_{2}(e_1\nu_1)\cdot{\sf L}_{2}(e_2\nu_2) {\sf L}_{0}^*(e_1\nu_i)
\cdot {\sf L}_{0}^*(e_2\nu_j)$ turns out to be identically null for
$i,j=1,2$ or $2,1$. Therefore,
{\em there is no interference term between the A and U matrix elements.},
as happens with ${\cal M}_{2\nu}^{\sss A}$ and
${\cal M}_{2\nu}^{\sss F}$. We get
\br
d\Gamma_{2\nu}&\equiv& d\Gamma_{2\nu}^{\sss A+NU} +d\Gamma_{2\nu}^{\sss U}
= \frac{4G^4}{15\pi^5} \left[ \left|{\cal M}_{2\nu}^{\sss A}
+{\cal M}_{2\nu}^{\sss NU}\right|^2 d\Omega_{2\nu}^{\sss A}+
|{\cal M}_{2\nu}^{\sss U}|^2 d\Omega_{2\nu}^{\sss U}\right],
\label{10}\er
where
\be
d\Omega_{2\nu}^{\sss A}
=\frac{1}{2^6\pi^2}(Q-\epsilon_1-\epsilon_2)^5
\prod_{k=1}^2p_k\epsilon_kF_0(\epsilon_k)d\epsilon_k,
\label{11}\ee
is the usual phase space for the $\beta\beta_{2\nu}$ in the A approximation,
and
\be
d\Omega_{2\nu}^{\sss U}= \frac{5^2}{2^{11}3^2\pi^2}
\sum_{i=0}^2a_i (Q - \epsilon_1- \epsilon_2)^{5+2i}
\prod_{k=1}^2p_k\epsilon_kF_0(\epsilon_k)d\epsilon_k,
\label{12}\ee
with $a_0={\bar p}_1^2{\bar p}_2^2$, $a_1=16{\bar p}_1^2/35$ and
$a_2=1/21$.

The  corresponding half life is
\be
T_{2\nu}(0_{\sss I}^+{\rightarrow} 0_{\sss NU}^+)
=\left({\cal G}_{2\nu}^{\sss A}
\left|{\cal M}_{2\nu}^{\sss A}+{\cal M}_{2\nu}^{\sss NU}\right|^2
+{\cal G}_{2\nu}^{\sss U}|{\cal M}_{2\nu}^{\sss U}|^2\right)^{-1},
\label{13}\ee
where
\be
{\cal G}_{2\nu}^{\sss A,U}=\frac{4G^4}{15\pi^5\ln 2}
\int d\Omega_{2\nu}^{\sss A,U},
\label{14}\ee
are the kinematical factors.

The spectrum shapes $d\Gamma_{2\nu}^{\sss A+NU}/d\epsilon$ and
$d\Gamma_{2\nu}^{\sss U}/d\epsilon$ for $^{76}Ge$ are confronted in
Fig. 1.
At variance with the single $\beta$ emission, {\em the spectrum for
the U double beta process deviates from the allowed shape in the
low-energy region, but not for $\epsilon\cong Q$}. Thus, independently of
magnitude of $\Gamma_{2\nu}^{\sss U}$, the virtual states $J^\pi=2^-$ will
never interfere with the detection of the $\beta\beta_{0\nu}$ events.
Their spectra, still, can overlap with those engendered by the
$\beta\beta_{\sss M}$ decays. This is illustrated in same figure
for the case of the CM model.

\begin{table}[h]
\begin{center}
\caption {Kinematical factors ${\cal G}_{2\nu}$, and
the nuclear matrix elements ${\cal M}_{2\nu}$
 evaluated within the QRPA formalism.}
\label{tab2}
\bigskip
\begin{tabular}{cccccc}
\hline
Nucleus&${\cal G}_{2\nu}^{\sss A}~[y^{-1}]$&${\cal G}_{2\nu}^{\sss U}
~[y^{-1}]$&${\cal M}_{2\nu}^{\sss A}$&${\cal M}_{2\nu}^{\sss NU}$
&${\cal M}_{2\nu}^{\sss U}$\\
\hline
$^{76}Ge$&$\!5.39~10^{-20}\!$&$\!2.10~10^{-19}\!$&$\!0.050\!$&$\!-0.008\!$
&$\!1.0~10^{-5}\!$\\
$^{82}Se$&$\!1.80~10^{-18}\!$&$\!2.54~10^{-17}\!$&$\!0.060\!$&$\!-0.009\!$
&$\!9.8~10^{-6}\!$\\
$^{100}Mo$&$\!3.91~10^{-18}\!$&$\!5.50~10^{-17}\!$&$\!0.051\!$&$\!-0.014\!$
&$\!1.1~10^{-5}\!$\\
\hline\end{tabular}\end{center}
\end{table}

Numerical results for the kinematical factors and the nuclear matrix elements,
for several experimentally interested nuclei, are displayed in Table
\ref{tab2}.
The moments ${\cal M}_{2\nu}$ were evaluated within the pn-QRPA model,
following the procedure
adopted in our previous works \cite{Bar95,Krm94}.
It can be easily seen that:
\[
|{\cal M}_{2\nu}^{\sss U}|\cong
R^2|{\cal M}_{2\nu}^{\sss A} +{\cal M}_{2\nu}^{\sss NU}|;~~~~
{\cal G}_{2\nu}^{\sss U}\cong
(Q/4)^4{\cal G}_{2\nu}^{\sss A}.
\]
Therefore, from the theoretical developments and numerical calculations
done here, it can be stated that
the simple estimate
\[T_{2\nu}^{\sss U}/
T_{2\nu}^{\sss A+NU}\cong (RQ/4)^{-4} \sim 10^{6},
\]
is appropriate for the unique transitions.

\begin{table}[h]
\begin{center}
\caption { Calculated half-lives (in units of y) for the A+NU and
 U transitions
and for the charge Majoron emission.}
\label{tab3}
\bigskip
\begin{tabular}{ccccc}
\hline
Nucleus&$T^{\sss A+F}_{2\nu}$&$T^{\sss U}_{2\nu}$
&$T_{\sss CM}({\ss M_+\rightarrow\infty})$&$T_{\sss CM}
({\ss M_+\cong M_-})$ \\
\hline
$^{76}Ge$  &$\!1.1~10^{22}\!$&$\!4.7~10^{28}\!$&$\!1.6~10^{25}\!$
&$\!1.3~10^{29}\!$\\
$^{82}Se$  &$\!2.1~10^{20}\!$&$\!4.1~10^{26}\!$&$\!9.2~10^{23}\!$
&$\!9.5~10^{27}\!$\\
$^{100}Mo$ &$\!1.9~10^{20}\!$&$\!1.4~10^{26}\!$&$\!3.8~10^{23}\!$
&$\!4.1~10^{27}\!$\\
\hline\end{tabular}\end{center}
\end{table}

Finally, in Table \ref{tab3} are compared the half-lives for the
$\beta\beta_{2\nu}$ and $\beta\beta_{\sss CM}$ decays.
In the CM model the effective coupling constant was taken to be
$g_{\sss CM}=\theta^2/2$ with $\theta=0.1$, and two different
values for the heavy Dirac neutrinos masses were considered,
namely $M_+\rightarrow \infty$ and $M=100$ MeV, and
$M_{\pm}=M\sqrt{1\pm \theta}$,
with  $M=100$ MeV \cite{Bur93,Bar96}.
It turns out that:
\[
T_{\sss CM}({\ss M_+\rightarrow\infty})/T_{2\nu}^{\sss A+NU}\sim 10^{3};~
T_{\sss CM}({\ss M_+\cong M_-})/T_{2\nu}^{\sss A+NU}\sim 10^7.\]
Thus, the emission rate for the recently discovered Majoron models
is very strongly conditioned
by the model parameters, and it could be so small
as that arising from the unique forbidden
transitions.

In summary, the higher order effects in the standard physics modify the
$\beta\beta_{2\nu}$ spectrum shape but only at the level of $10^{-6}$
and mainly at low two-electron energy, where most backgrounds tend to dominate.
Therefore they would hardly mask the observation of the potential exotic
$\beta\beta$ decays.


\newpage
{\bf Figure Captions}

\vskip 0.5cm

Fig. 1. Electron energy spectrum for the nucleus $^{76}Ge$, as a function
of the sum of energies of the two emitted electrons, for: the standard $2\nu$
allowed ($\beta\beta_{2\nu}^A$) and unique forbidden transitions
($\beta\beta_{2\nu}^{U}$), and the exotic neutrinoless decays, with Majoron
charged emission ($\beta\beta_{\sss CM}$) and without ($\beta\beta_{0\nu}$).
All four curves have been arbitrarily assigned the same maximal values
for purposes of comparison.
\end{document}